| | |
|---|---|
| **Manuscript title:** | Improving the diagnosis of breast cancer based on biophysical ultrasound features utilizing machine learning |
| **Authors' names:** | Jihye Baek[1], Avice M. O'Connell[2], Kevin J. Parker[1] |
| **Institutional affiliations:** | [1]Department of Electrical and Computer Engineering, University of Rochester, Rochester, NY USA, [2]Department of Imaging Sciences, University of Rochester Medical Center, Rochester, USA |
| **Corresponding author:** | Kevin J. Parker<br>University of Rochester<br>Computer Studies Building 724<br>Box 270231<br>Rochester, NY 14627-0231<br>Phone: (585)275-3294<br>Fax: (585)275-2073<br>Email: kevin.parker@rochester.edu |

## Abstract

The improved diagnostic accuracy of ultrasound breast examinations remains an important goal. In this study, we propose a biophysical feature-based machine learning method for breast cancer detection to improve the performance beyond a benchmark deep learning algorithm and to furthermore provide a color overlay visual map of the probability of malignancy within a lesion. This overall framework is termed disease specific imaging. Previously, 150 breast lesions were segmented and classified utilizing a modified fully convolutional network and a modified GoogLeNet, respectively. In this study multiparametric analysis was performed within the contoured lesions. Features were extracted from ultrasound radiofrequency, envelope, and log-compressed data based on biophysical and morphological models. The support vector machine with a Gaussian kernel constructed a nonlinear hyperplane, and we calculated the distance between the hyperplane and each feature's data point in multiparametric space. The distance can quantitatively assess a lesion, and suggest the probability of malignancy that is color-coded and overlaid onto B-mode images. Training and evaluation were performed on in vivo patient data. The overall accuracy for the most common types and sizes of breast lesions in our study exceeded 98.0% for classification and 0.98 for an area under the receiver operating characteristic curve, which is more precise than the performance of radiologists and a deep learning system. Further, the correlation between the probability and BI-RADS enables a quantitative guideline to predict breast cancer. Therefore, we anticipate that the proposed framework can help radiologists achieve more accurate and convenient breast cancer classification and detection.

# I. INTRODUCTION

There are a growing number of approaches that incorporate machine learning into the diagnosis of breast cancer using ultrasound and other imaging modalities [1-4]. Within these approaches, ultrasound imaging has several advantages. Ultrasound is relatively affordable with a lower price point among the portable units. This makes ultrasound a widely accessible, non-ionizing method for imaging studies, including underserved populations in developing countries[5-7]. Breast ultrasound also is utilized as an adjunct to x-ray mammography in certain cases, particularly the dense breast. Given these advantages, an intensive effort has been made to improve breast ultrasound using computer-assisted analyses over recent decades. Earlier approaches incorporated lesions' features such as size, shape, texture, and boundaries within clustering and classification systems or rule-based algorithms[8-11].

In recent years, further developments in artificial intelligence (AI) have broadened the types of breast cancer analyses. Previous studies utilizing machine learning approaches, such as support vector machine[12-14] and random forest [15], have output breast classifications of benign or malignant. These typically include feature extraction and selection, and their performance and running time rely on the efficiency of this step [11]. Thus, machine learning approaches for breasts traditionally extracted simple texture or morphological features from log-compressed B-mode images. However, how to optimize feature extraction and selection remains unclear yet critical for performance. To address this feature dependency on performance and time-consuming process, deep learning algorithms [16-22] have been recently applied to breast cancer detection not only in classification but also in lesion segmentation. Deep learning can advantageously extract data-driven and self-optimized feature maps from input images, and thus the feature detection and selection are unnecessary. However, many approaches incorporate large training sets to produce accurate diagnostic classifications, but collecting a large number of patient ultrasound data is challenging [1, 11, 23] . Moreover, due to the computational complexity of deep learning, there are limitations to the size and type of ultrasound signal inputs. Specifically, input data after postprocessing, including log-compression and speckle reduction, are commonly used. The raw ultrasound signals are not utilized, which limits classification performance because the input has lost much information compared to raw ultrasound signals. Therefore, new approaches to using the information on raw ultrasound signals while exploiting the advantages of deep learning and machine learning would be valuable.

A recent study [24] attempted a deep learning approach to predict breast treatment response using the quantitative ultrasound (QUS) multiparametric maps, yielding comparable performance with previous deep learning studies. This study extracted features from raw ultrasound signals and added this information as "preprocessing". The extracted QUS features were used as input to the algorithm, and then the algorithm extracted QUS map-driven features once again and predicted two outputs of responder or non-responder to therapy. In contrast, we utilize raw ultrasound signal-driven features as "postprocessing" of deep learning for improved diagnosis. Furthermore, beyond the two outputs of deep learning (benign/malignant), we designed our approach to provide probability of malignancy utilizing the features and machine learning: our features are grounded on biophysical models of ultrasound-tissue interactions [25-28].

In this paper, our goal is to improve the diagnostic capabilities of breast lesions imaged using conventional ultrasound scanners. To do this, we combine a number of recently developed biophysics-based analyses of ultrasound echoes, specifically the H-scan analysis and Burr analysis of speckle [29, 30] , along with additional morphological measures of the lesion. These are com-bined mathematically using principal component analysis (PCA) and are separated in a multiparametric space by the support vector machine (SVM). The results of this synthesis, applied to a curated data base of 157 breast scans, demonstrate higher area under the curve (AUC), sensitivity, and specificity than has been achieved in earlier studies, including multiparametric analysis and AI studies. In addition to this approach, the overall analysis enables a visual display of the multiparametric and classification analyses, whereby colored overlays on the breast lesion indicate the localized probability of malignancy. This feature, known as disease-specific imaging (DSI), enables an immediate visual "gestalt" that is grounded in quantitative analysis and SVM classification metrics.

## II. THEORY

*A. Probability of malignancy and disease-specific imaging for breast cancer*

DSI is an imaging approach which utilizes a multiparametric analysis to classify disease types and visualize disease progression and severity. The DSI approach first utilizes SVM for classification of disease categories within a training set and then assesses disease severity in any individual patient by multiparametric measures within the classified, multidimensional space. Lastly, the disease types and progression levels are assigned unique colors and color intensities overlaid on B-mode images. The DSI framework has been applied to liver diseases, where the SVM can classify liver fibrosis, steatosis, cancers, and normal liver. Once a disease category of a patient is specified, then its progression level is estimated utilizing the inner product operation within the multidimensional, multiparametric coordinate system [31].

However, in the area of ultrasound breast exams, lesions are generally categorized as benign and malignant, and there are many sub-types within these categories. Some lesions within benign and malignant categories may have only slightly different multiparametric measures, and so the disease trajectories would have overlaps. In this paper, we consider these issues in breast cancer studies. We propose a newly modified DSI (Fig. 1), which skips the classification step of lesion sub-types but utilizes the SVM for disease progression quantification, including the classification of benign and malignant. We assigned a common color map for breast lesions, where color levels are distributed from green to red, indicating the likelihood of benign or malignant lesions, respectively. To enable this approach, the SVM builds a non-linear hyperplane classifying benign and malignant classes, and for any individual lesion the distance from the hyperplane is used for quantification of breast disease progression, indicating the probability of malignancy. The details of this approach are described in the next Sections.

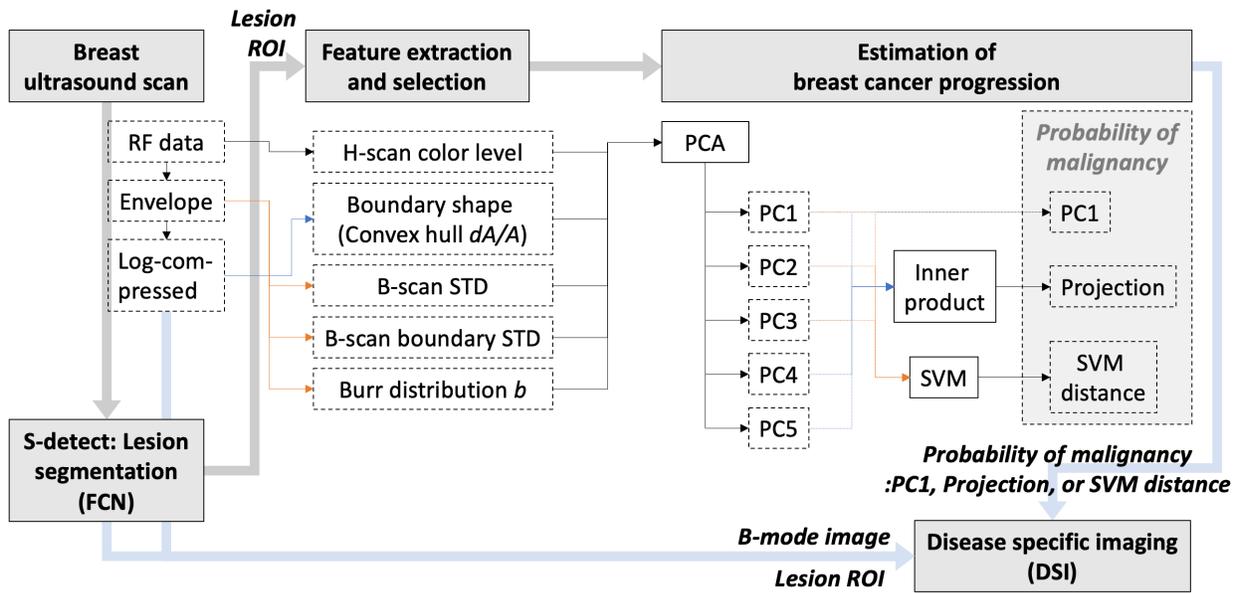

**Figure 1** Flow chart for our breast cancer identification, including lesion detection, estimation of cancer progression, and imaging.

*B. Quantification of Breast Cancer Progression*

Many parameters characterizing tissue signatures have been introduced, for example, several features extracted using ultrasound speckle, attenuation from beam propagation, shear wave attenuation and speed, and parameters extracted from frequency domain analyses including the H-scan [25, 27, 28, 31, 32]. Multiparametric analysis can be performed to combine more information from the multiple parameters while excluding their dependencies. Principal component analysis is a simple method to combine the parameters. The first principal component (PC1) tends to include most efficient independent information from the parameters, and therefore in some cases PC1 by itself can be simply used for estimating disease progression levels. The PC1 for DSI is a simple and fast approach to implement DSI for clinical ultrasound systems. However, it loses information from the second and higher principal components, and thus for more accurate estimation, a procedure to include several other principal components may be advantageous.

As proposed in our previous work [31, 33], the inner product resulting in projection can compare the similarity of multiple parameters to a reference set of measures associated with a particular disease. This assumes linear trajectories and simple disease model, and was found to be effective on animal models. In this study we generalize the concept for distinction between benign and malignant lesions without requiring unique labels of specific sub-types.

*C. SVM Classification distance*

To construct a hyperplane for differentiation between benign and malignant, SVM training is performed with ground truth biopsy results. The use of a Gaussian kernel for the SVM allows us to develop a non-linear hyperplane. The SVM parameters of the box constant and kernel parameter are optimized to obtain a smooth and robust hyperplane, avoiding overfitting. More details regarding SVM parameter optimization can be found in [27]. Once we define a non-linear hyperplane, we can calculate a distance between the plane and any data point in the multiparametric space (representing measurement from an individual lesion) to quantify disease progression, denoting the probability of malignancy, as illustrated in Fig. 2 (a). This SVM distance is defined by:

$$sign \cdot Distance_{feature-SVM} \qquad (eq\ 1)$$

where $Distance_{feature-SVM}$ is an absolute distance between a measured feature and the hyperplane, and sign denotes classification outputs -1 and 1 for benign and malignant, respectively. As shown in Fig. 2 (b), we can set a lower and upper limit of the SVM distance, and the parameters have corresponding colors distributed from green to red, representing high probability of benign and malignant breast tissues, respectively. More details for the correlation between the color levels and probability are discussed in the section, IV. C. DSI Performance.

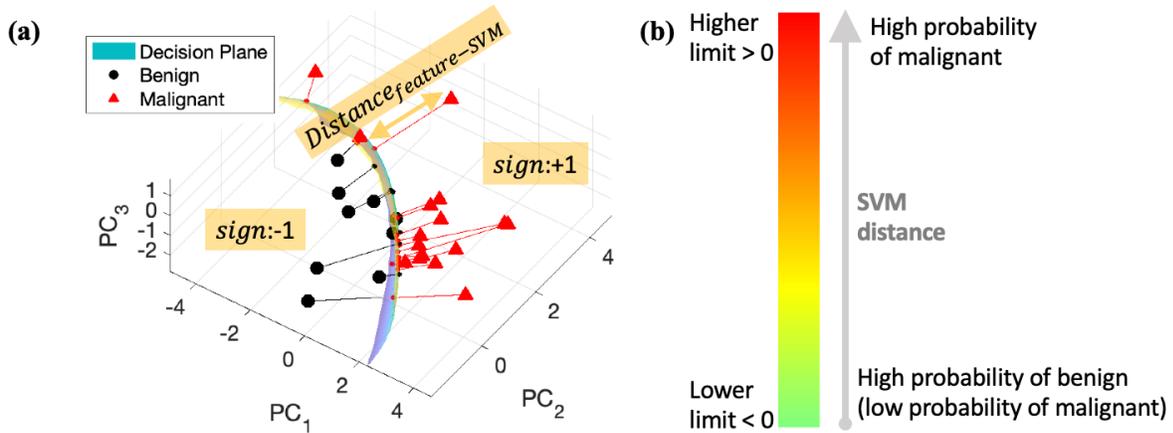

**Figure 2** SVM distance and color map. (a) An example SVM hyperplane with measured features are shown. The area classified as benign and malignant have *sign* -1 and +1, respectively. The distance between a feature and the plane is $Distance_{feature-SVM}$; an example distance is indicated by the orange arrow. (b) SVM distance mapping to a color map, representing probability of benign and malignant.

### III. METHODS

*A. Study protocol*

This study was approved by the requirements of informed consent of the Research Subjects Review Board at the University of Rochester. We enrolled 157 patients at the University of Rochester Medical Center, who have at least one suspicious breast lesion and were recommended to undergo biopsy or ultrasound imaging follow-up. The biopsy results were considered as "gold standard." Patient characteristics are provided in Table I; among the 157 patients, 7 patients were excluded because the study protocol was not followed correctly. An experienced radiologist also categorized the breast lesion diagnoses into three groups: (1) major types (seen frequently), (2) common, and (3) uncommon cases (infrequently seen or rare).

Ultrasound examinations were performed using the RS85 ultrasound scanner (Samsung Medison Co., Ltd., South Korea) and a 3–12 MHz linear array transducer (L3-12A). A static image of the suspicious mass/lesion was acquired for each patient after the first screening of the lesion in the longitudinal and transverse planes. Among the 157 patients, only 154 patient scans were saved as radiofrequency (RF) data; for the others, only cine loops (without RF data) were saved, which cannot be analyzed for this study. The RF data were sent to Samsung Medison Co., Inc. (Seoul, South Korea), where they were converted into in-phase and quadrature (IQ) data (proprietary process for the RF data and its header information). We were able to reconstruct RF data based on scan information provided by the company. From this reconstructed data, only 121 patient scans which met the following consistent conditions were included in this study:

- patients for whom the study protocol was followed correctly.
- a 9.4 MHz center frequency for transmission.
- without using harmonic mode.
- with BI-RADS scores provided by radiologists.

The ultrasound breast images were reviewed by ten board-certified or board-eligible radiologists who have less than 5 years (n = 5) or over 10 years (n = 5) of experience. Thus, the only IQ data included were those where radiologists provided BI-RADS scores.

The BI-RADS scores provided by the 10 radiologists were averaged by an area-preserving method of receiver operating characteristic curve averaging [34]. The averaged BI-RADS score indicates the radiologists' performance in breast diagnosis. Further, a deep learning framework to classify breast lesions, called S-detect in this paper, (S-DetectTM for Breast in RS80A, Samsung Medison Co., Ltd., Seoul, Korea) [35, 36], was utilized, which indicated the performance of deep learning; more details are found in the section III. B. Breast Lesion Contour Utilizing Deep Learning. S-detect classifies breast lesions as benign or malignant and outputs lesion boundaries. Lastly, our proposed analysis was applied to ultrasound breast images. The proposed method can combine multiple parameters, provide a quantified parameter indicating the probability of malignancy, and show a visual display called DSI. To define a breast lesion in an ultrasound image, we used the S-detect output of a lesion boundary. Also, we approximated (or smoothed) boundaries based on the S-detect results for efficient processing for parameter estimations. The S-detect boundary was used for estimating boundary shape measured using the convex hull estimation described in the section III. C. Feature Extraction – Ultrasound Parameters 2). The other parameters were not sensitive to boundary shapes, and thus approximated boundaries were used.

| Characteristic | Categories(count) / mean $\pm$ standard deviation |
|---|---|
| Age | 52.75 $\pm$ 14.89 |
| Ethic origin | White (n = 106), African American (n = 33), |

|  | Asian (n = 3), |
|  | Other categories (n = 8) |
| Lesion area size | 0.87±1.09 cm$^2$ |
| Histology | Benign (n = 94), |
|  | Malignant (n = 56) |
| Breast density | Almost entirely fatty (n = 9), |
|  | Scattered areas of fibro glandular density (n = 41), |
|  | Heterogeneously dense (n = 78), |
|  | Extremely dense (n = 15), |
|  | Density not available (n = 7) |
| List of Major Categories | Major Benign (n = 68), |
|  | Major Malignant (n = 32), |
|  | Common (n = 46, benign = 26/malignant = 20), |
|  | Uncommon (n = 4) |
| Major Benign | Stromal fibrosis |
|  | Fibroadenoma |
|  | Fibro adenomatoid changes (FAC) |
|  | Fibrocystic changes with stromal fibrosis Intraductal papilloma |
|  | Cyst (microcyst cluster, ruptured cyst, simple cyst) |
|  | Follow-up stable mass |
| Major Malignant | Ductal carcinoma in situ (DCIS) |
|  | Invasive ductal carcinoma (IDC) |
|  | Invasive lobular carcinoma (ILC) |

| | Invasive ductal and lobular carcinoma Invasive ductal carcinoma with micropapillary features |
| | Invasive mammary carcinoma (IMC) |

**Table 1** Patient information

*B. Breast lesion contour utilizing deep learning*

A breast lesion in an ultrasound image was contoured by the S-detect. The S-detect was developed as a deep learning-based breast diagnosis system for ultrasound images, and classifies breast lesions as benign or malignant.

First, a breast lesion boundary is segmented by a modified fully convolutional network (FCN). Although the FCN [37] can automatically segment a lesion, S-detect modified the FCN for semi-automatic segmentation to reduce segmentation error due to unclear lesions on ultrasound images. Thus, the modified FCN requires clinicians to specify the center of a lesion and then segments a boundary near the center. The next step is classification of breast tissues within the lesion boundary. GoogLeNet [38] was modified by removing two auxiliary classifiers. Further, its input layer was modified to use gray scale ultrasound images rather than color input images, and it has two class outputs of benign or malignant rather than 1000 classes in GoogLeNet.

S-detect utilizes log-compressed envelope data as its input images. The input image is resized to a fixed pixel × pixel 2D image, and therefore each image with a lesion has different resize rate. Larger lesions have higher down-sampling rate than smaller lesions, causing more information loss for the larger lesions.

The S-detect was pretrained [36] and used in this study. The training was performed with 790 images acquired using the RS80A system (Samsung Medison Co., Ltd.) and 6775 images acquired using other ultrasound machines (iU22, Philips Healthcare, Bothell, WA, USA; A30, Samsung

Medison Co., Ltd.). S-detect performance identifying breast lesions was reported in a previous study [35].

*C. Feature extraction – Ultrasound parameters*

We extracted features from ultrasound signals. Reflected ultrasound echoes are converted from RF data to IQ data, envelope data, and log-compressed data, in sequence. Deep learning approaches generally extract features from log-compressed data. However, to obtain higher accuracy based on multiparametric analysis rather than utilizing deep learning, we also extracted features from previous data, including RF data and envelope data, since the previous data have more information than the log-compressed data. For example, we can extract frequency information from RF data, but it cannot be found from envelope or log-compressed envelope data. Our selected features are presented in Fig. 1 with input data.

1) H-scan color level

The H-scan is a matched filter analysis [29] , which enables the extraction of frequency information, including frequency spectrum distribution and shift, caused by attenuation and medium/scatterer changes in size or connections between scatterers. H-scan processing and example results are shown in Fig. 3. The frequency at each time sample can be estimated by matched filters. This study used 256 Gaussian functions working as bandpass filters in the frequency domain; each Gaussian has its peak frequency from 5.2 MHz to 12.4 MHz in an equal frequency difference of $2.8 \times 10^{-1}$ MHz. The filtering output using ultrasound RF data and the Gaussian filters resulted in 256 convolutional images. By selecting a maximum convolutional value at each time sample and each scanline, we can obtain the corresponding Gaussian index ($i_{MAX}(t)$ in Figure 3) with a peak frequency for the maximum. The Gaussian index ranging from 1 to 256 becomes a color level ($C \in \{1, 2, \ldots, 256\}$ in Figure 3); the lower color levels indicate lower frequency components, while the higher levels indicate higher frequency components. The lower and higher components can denote larger and smaller scatterers, respectively. And the lower to higher color levels are mapped into the red to blue colors of the H-scan; the H-scan color bar is shown in Fig. 3. In this study as a breast application, the red to blue colors represent lower and higher probability of malignancy (higher and lower probability of benign), respectively.

Ultrasound propagation along with depth causes attenuation, resulting in a red-shift of H-scan colors over depth. To compensate for this attenuation effect, attenuation-corrected RF data were used as inputs of the H-scan. For this correction [29], we applied $e^{\alpha f_0 x_z}$ to frequency spectrum $S_z(f)$ of RF data divided by 10 depth zones over depth, where $\alpha$ is attenuation coefficient, $f_0$ is center frequency, $x_z$ is average depth of each zone $z$ from 1 to 10, and $S_z(f)$ is frequency spectrum of zone $z$. We used $f_0$ = 9.4 MHz and $\alpha$ = 1 dB/MHz/cm for our breast data. The measured color levels within a breast lesion for a patient were used as an H-scan parameter ranging from 1 to 256, called the "H-scan color level."

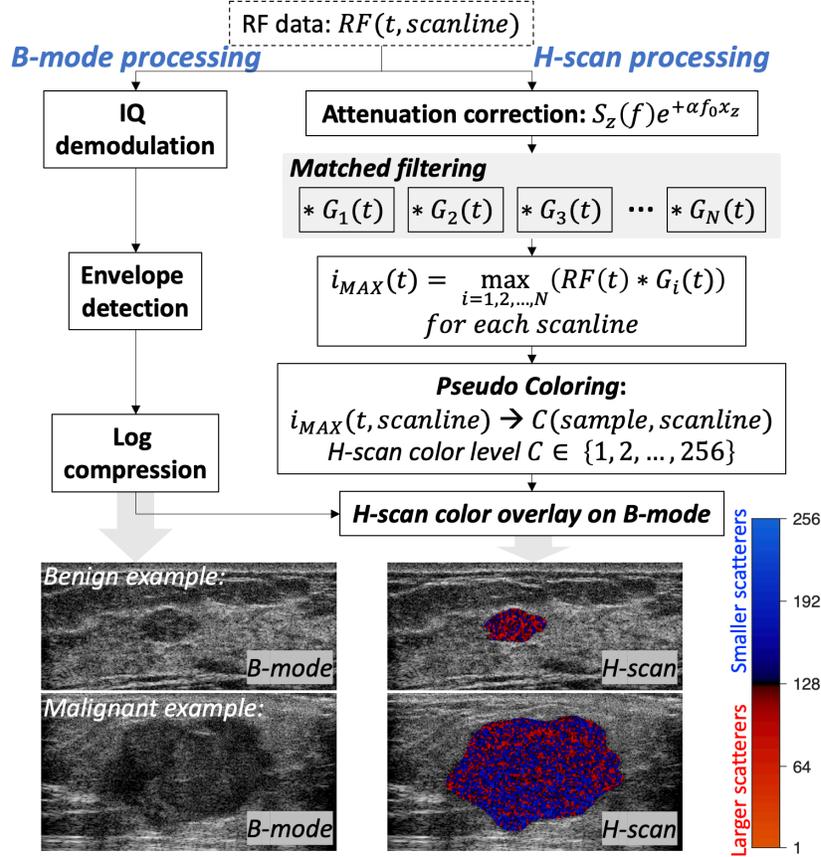

**Figure 3** H-scan processing and example results of benign and malignant cases.

2) Boundary shape measured using the convex hull estimation [dA/A]

Breast margin shape is one of the most critical features in BI-RADS scoring to classify breast lesions as benign or malignant, since benign lesions generally show smoother boundary shape than malignant lesions. In previous research, morphologic features were introduced and utilized to delineate breast lesion boundaries, utilizing ellipse fit, convex hull, concave hull, and roundness [8, 11] . In this study, we employed convex hull due to its relatively lower computation time. We have also tested ellipse fit as well as the convex hull estimation. Finding an optimal fit after several trials of ellipse fits is more time-consuming than finding a unique convex hull for any contours, but the boundary estimation accuracy was comparable for both methods. Therefore, with the convex hull estimation, we proposed a calculation method to accurately measure boundary roughness. As shown in Fig. 4, convex hull of a breast lesion was obtained, and we estimated the lesion shape roughness using (eq 2):

$$\frac{dA}{A} = \frac{(ConvexhullArea) - (ContouredArea)}{(ContouredArea)} \quad \text{(eq 2)}$$

where $dA$ is the area difference between the convex hull area and the contoured area, and $A$ is the contoured area of a lesion. The boundary shape was defined by the proportion of the area difference ($dA$) to the lesion size $A$. The smoother boundaries have smaller $dA/A$, and thus this can a feature to estimate the boundary shape of breast lesions.

To calculate this feature, log-compressed envelope data were used as input ultrasound images, and lesion contour using S-detect was utilized as input of the boundary shape measurement.

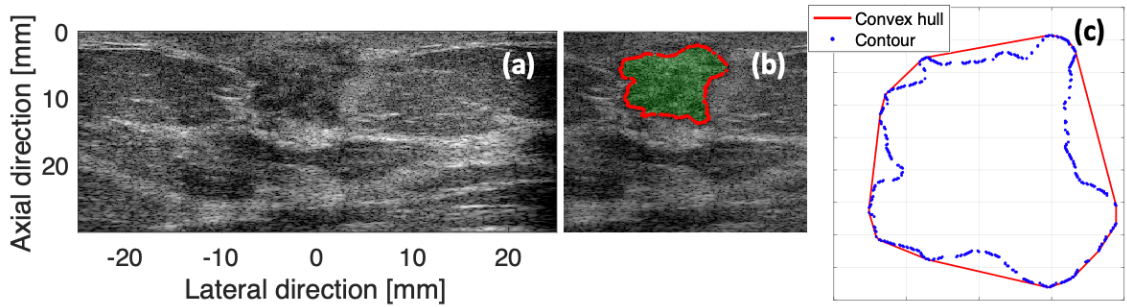

**Figure 4** Boundary shape measured by convex hull. (a) An example breast scan. (b) Contoured lesion using a deep learning approach. (c) Convex hull of the lesion. Using the area within the contour (A) and the area difference between the contour and convex hulls (dA), boundary shape roughness of a breast lesion (dA/A) is measured.

3) B-scan parameters

To extract texture features, we simply calculated B-scan intensity and standard deviation (STD) using log-compressed envelope data within a lesion to detect mathematical patterns.

4) B-scan boundary parameters

In addition to extracting features within a lesion, we also detected a texture feature utilizing information near the lesion boundary. Characteristics of normal breast tissues and tissues in a lesion including benign and malignant tumors are different, and thus the normal and abnormal tissues tend to be differentiated in B-scan images or other parametric ultrasound images. Fig. 5 (a) shows an example breast image with a margin, and we can find the boundary between normal and abnormal tissues. Based on the boundary, the lesion was highlighted in green as shown in Fig. 5 (b). Depending on the lesion type, the boundaries may be more or less clear, and this boundary characteristic can help classification of breast conditions. Therefore, we investigated tissues near the lesion boundary including inner and outer margins, as shown in Fig. 5 (e-f), and within the margin, we extracted features using B-scan imaging: B-scan boundary intensity and STD. The inner and outer margins were defined using morphological erosion and dilation, respectively. Using the green lesion in Fig. 5 (c-d), erosion and dilation generated the blue and red area, respectively. The green in (c) and red in (d) became the inner and outer margins of the lesion, respectively, which is shown in Fig. 5 (e). Both margins combined are the boundary area as shown in Fig. 5 (f). The morphological operations were performed with a disk shown in Fig. 5 (e), and the disk size was determined so as to set the margin length at 10% of the lesion length. Then, within the margin in Fig. 5 (f), we calculated B-scan intensity and STD.

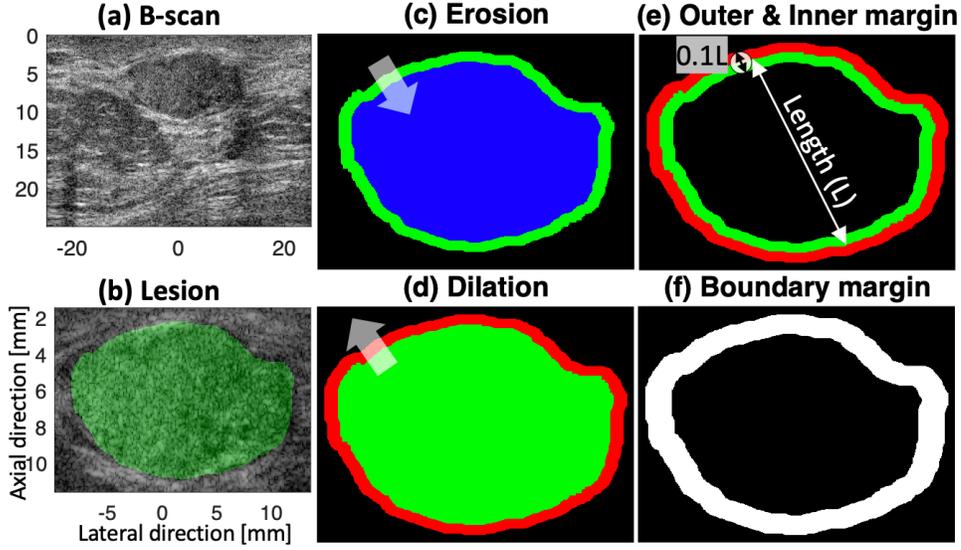

**Figure 5** (a) An example B-scan image. (b) A lesion highlighted in green. (c) Morphological erosion was used to generate the blue area from the green lesion. (d) Morphological dilation was used to generate the red area from the green lesion. (e) The morphological erosion and dilation resulted in the inner and outer margins highlighted in green and red, respectively. (f) The margins in (e) were used as a boundary margin to calculate the feature "B-scan boundary standard deviation (STD)."

5) Burr power law parameters

We observed the histogram of ultrasound envelope data, and previous studies [39, 40] revealed that the histogram distribution of many soft tissuses as a probability density function $P(A)$ is governed by the Burr distribution:

$$P(A) = \frac{2A(b-1)}{\lambda^2 \left[\left(\frac{A}{\lambda}\right)^2 + 1\right]^b}$$

(eq 3)

where *A* is an echo amplitude, and the distribution has two parameters of $\lambda$ (a scale factor that increases with amplitude and gain) and *b* (a power law exponent dependent on scatterer distributions). Using an ultrasound envelope data, the curve fitting using the Burr distribution estimates the two parameters that can be used as features of our multiparametric analysis.

One practical matter regarding this histogram estimation is the optimal setting of the histogram bin number as a sampling rate of the echo amplitudes distributed as a continuous real number. The sampling rate can be coarse, but it should be able to describe specific shapes of histogram distributions. Further, it should be sufficient enough to provide information enabling the curve fitting. However, an optimal sampling rate for each ultrasound frame can be different, and therefore we calculated the *R*-squared ($R^2$) for the Burr curve fitting within the sampling rates between 2% and 40% with equal intervals of 2%; the observed percentages are 2%, 4%, 6%, …, 38%, and 40%. As illustrated in Fig. 6, $R^2$ for the percentages were observed. Figure 6 (a) shows an example B-scan, and Figure 6 (b) shows a lesion in the images. Envelope data within the lesion were utilized for the curve fittings with different sampling rates. However, the estimated $\lambda$ and *b* tended not to vary depending on the sampling rates; the standard deviations of $\lambda$ and *b* estimation are 0.0287 and 0.0015, respectively. $R^2$ tended to decrease with increasing the sampling percentages. Therefore, we selected the sampling rate of 10% to have $R^2$ over 0.96 and to include enough bin numbers. We also considered the histogram bin numbers due to small lesion sizes, and the minimum bin number for the 10% is 229; but the minimum bin number for 2% is only 46, which may be insufficient for fitting. In conclusion, the sampling rate of 10% was founded as an optimal for all patient data, and was used to find optimal Burr parameters of $\lambda$ and *b*.

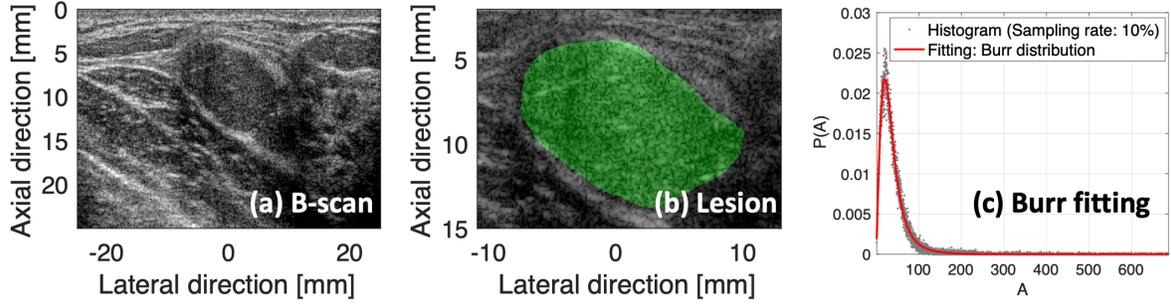

**Figure 6** Burr distribution estimation of an example breast scan. (a) and (b) show an B-mode image and a lesion in the image, respectively. (c) Example histogram plotted and Burr-fitted distribution.

*D. Feature selection*

As mentioned in the previous section, Feature Extraction, we estimated the following ultrasound parameters: H-scan color level and STD, boundary shape parameter, B-scan parameters of intensity and STD, B-scan boundary parameters of intensity and STD, and Burr power law parameters of $\lambda$ and $b$. Using the ten parameters, we have selected five: H-scan color level, boundary shape parameter, B-scan STD, B-scan boundary STD, and Burr power law. For the feature selection, AUC scores were compared for different parameter combinations. Receiver operating characteristic (ROC) curves were obtained using (1) the ground truth of biopsy results (benign or malignant), and (2) the performance observed from each parameter combination, which was measured by PC1 and SVM distance as the probability of malignancy. The AUC for the 5-parameter combination was the highest compared to any other parameter combinations, and therefore those features were selected for our analysis. The H-scan color level is a local parameter, whereas the other four are global parameters.

*E. Evaluation metrics*

To assess the accuracy of our proposed methods, we utilized the following evaluation metrics: (1) AUC, accuracy, sensitivity, and specificity from ROC curves, and (2) classification accuracy using the SVM. The metrics were obtained for our quantified outputs of PC1, projection, and SVM distance. Further, these outputs were compared with doctors' BI-RADS scores to compare our results with doctors' diagnostic performance.

To investigate lesion size dependance, we evaluated the performance of smaller to larger lesions, where lesion size thresholds of 0 $cm^2$, 0.1 $cm^2$, …, 0.9 $cm^2$, 1.0 $cm^2$ were observed. To be specific, for a threshold A $cm^2$, only breast lesions greater than A $cm^2$ were included for performance evaluations.

Moreover, we investigated the performance for all cases and for major benign and malignant. We first grouped the enrolled patients into (1) major benign and malignant, (2) common, and (3) uncommon cases. The all case group includes all enrolled patients, but major benign/malignant only includes major benign and malignant cases, excluding the common and uncommon cases.

DSI provides qualitative and quantitative results, which are the imaging outputs and probability of malignancy (suggested by PC1, projection, or SVM distance), respectively. The DSI requires training for breast condition prediction. Therefore, we divided the enrolled cases into training and testing sets, and the previously mentioned evaluation metrics for the testing set, in addition to the training set, were observed to examine the potential prediction precision.

*F. Training and Testing Data Sets*

Since the total number of data in this study (n = 121) is insufficient to divide into training and testing sets, we generated 5 combinations of training/testing sets. Each training and testing set was randomly divided; the training and testing sets include 70% and 30% of data, respectively.

This random generation was repeated 5 times, resulting in 5 randomly generated training/testing sets. Our results including classification accuracy were averaged over the 5 sets.

## IV. RESULTS

### A. Features and Contributions

We extracted the ten features described in the Methods section. The five features of H-scan color level, boundary shape, B-scan STD, B-scan boundary STD, and Burr b were selected since this combination resulted in the highest AUC compared to other parameter combinations using the ten features. To obtain the AUC, we plotted ROC curves using PC1/projection/SVM distance and ground truth of benign and malignant. To assess whether the parameters can differentiate benign and malignant, Fig. 7 (a-e) show p-values for each feature and Fig. 7 (f-h) show p-values for the combined parameters of PC1, projection, and SVM distance. It is revealed that each feature can differentiate benign and malignant, showing p-values less than 0.05; however, the combined parameters tend to result in lower p-values than each feature, meaning combining information from the five features provided better separation between benign and malignant. Fig. 7 (i) shows ROC curves for the five features and the three combined parameters, and Table II provides AUC values for the ROC curves. The five features showed relatively comparable AUC, but the three combined parameters resulted in higher AUC than each feature. Moreover, SVM distance resulted in the highest AUC, showing the best performance in separating the benign and malignant cases. Note that Fig. 7 includes all enrolled patient categories of major benign/malignant, common, and uncommon. AUC scores are presented in Table II. The 'All' column includes all categories. The 'Major' (benign and malignant) columns only included major cases after excluding common/uncommon to calculate the AUC scores. Investigating only major categories with lesion sizes greater than 0.4 cm$^2$ resulted in the highest AUC scores.

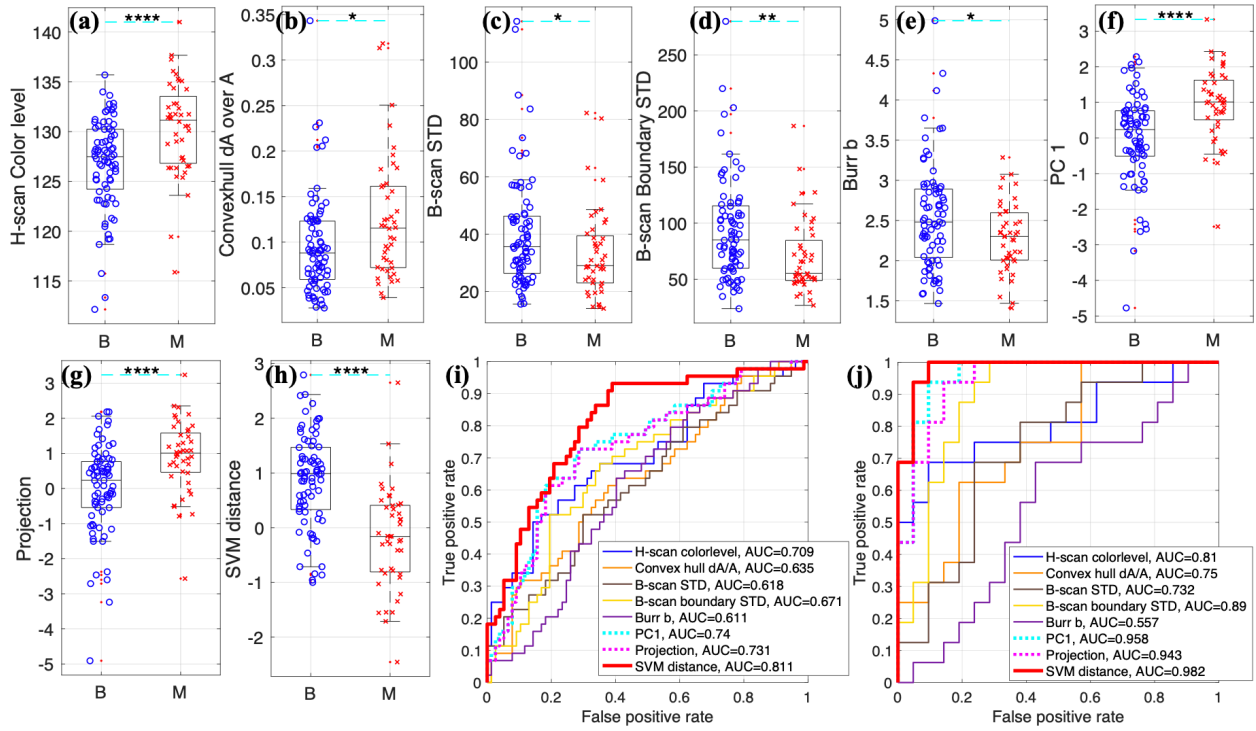

**Figure 7** Performance of each feature (a-e) and combined features (f-h): all categories were included. (i) ROC curves for all categories. (j) ROC curves for major categories with size over 0.4cm$^2$. The following notations are used for the statistics: ns (no significance) p > 0.05; * p < 0.05; ** p < 0.01; *** p < 0.001; and **** p < 0.0001. *GT: ground truth obtained from biopsy: B = Benign and M = Malignant

|  | Feature | AUC | | |
|---|---|---|---|---|
|  |  | All[a] (A > 0 cm$^2$) | Major[b] (A > 0 cm$^2$) | Major[c] (A > 0.4 cm$^2$) |
| Individual Feature | H-scan color level | 0.709 | 0.746 | 0.810 |
|  | Boundary shape using convex hull dA/A | 0.635 | 0.696 | 0.750 |

|  | B-scan STD | 0.618 | 0.650 | 0.732 |
|---|---|---|---|---|
|  | B-scan boundary STD | 0.671 | 0.751 | 0.890 |
|  | Burr $b$ | 0.611 | 0.614 | 0.557 |
| Combined Feature | PC1 | 0.740 | 0.866 | 0.958 |
|  | Projection | 0.731 | 0.860 | 0.943 |
|  | SVM distance | 0.811 | 0.866 | 0.982 |

**Table 3** All cases, all categories, and all lesion sizes were included to obtain the AUC scores in "All". Further, after excluding common/uncommon cases, only major benign/malignant cases were utilized to obtain the AUC scores in the right two columns with all lesion sizes and lesion sizes over 0.4 cm$^2$. However, for further analysis, specific sizes were also considered, which enables higher AUCs as will be shown in the following sections. [a]All: all lesion categories; [b]Major: major lesion categories

Our 3 combined parameters showed AUC greater than 0.86. These results demonstrate that we successfully selected features to combine information from each feature and to suggest a single combined parameter, and the best way to combine the parameters is the SVM distance. To investigate each feature's contribution for the combination, we utilized PCA, which is shown in Fig. 8. Using the five features, PCA was applied to obtain the principal components, and the weight factors to calculate principal components from each feature were summed and used to calculate their contribution in Fig. 8. The contributions for all cases and major categories are shown. H-scan and convex hull parameters tend to show higher contribution than the others. Boundary shape tends to make the highest contribution for the smaller lesion sizes, whereas H-scan tends to make the highest contribution for the larger lesion sizes. Further, the contributions from B-scan STD and Burr, which were based on ultrasound B-mode intensity, are likely to decrease as lesion sizes increase. As malignant lesions commonly have larger sizes and lower intensities, the hypoechoic area in larger sizes may provide less meaningful information than other features to distinguish tissue conditions, yielding less contribution than other features.

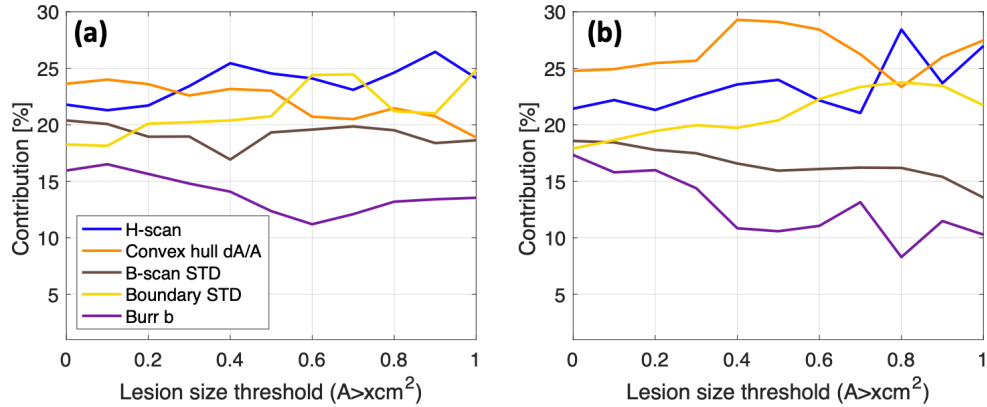

**Figure 8** Contribution from each feature. (a) Includes all categories of major benign/malignant (B/M), common, uncommon cases. (b) Includes only major benign and malignant cases.

*B. SVM classification and SVM distance*

The SVM classifier was used for calculating the SVM distance parameter and classifying benign and malignant cases. Fig. 9 (a-b) shows examples of SVM hyperplanes in two different views for training and testing sets. It also shows how we calculated the SVM distance: the SVM distance was calculated using the distance from each data point to the hyperplane, where a line corresponding to the distance is provided. Fig. 9 (c) shows classification accuracy for training and testing sets and also for all cases and major cases. The classification accuracy for major benign and malignant cases tends to be higher than all cases including major, common, and uncommon cases. The accuracy for the training sets is 80% to 100%, and that for the testing sets is approximately 70% to 80% for all cases and 80% to 90% for major cases. The larger lesion sizes resulted in higher classification accuracies.

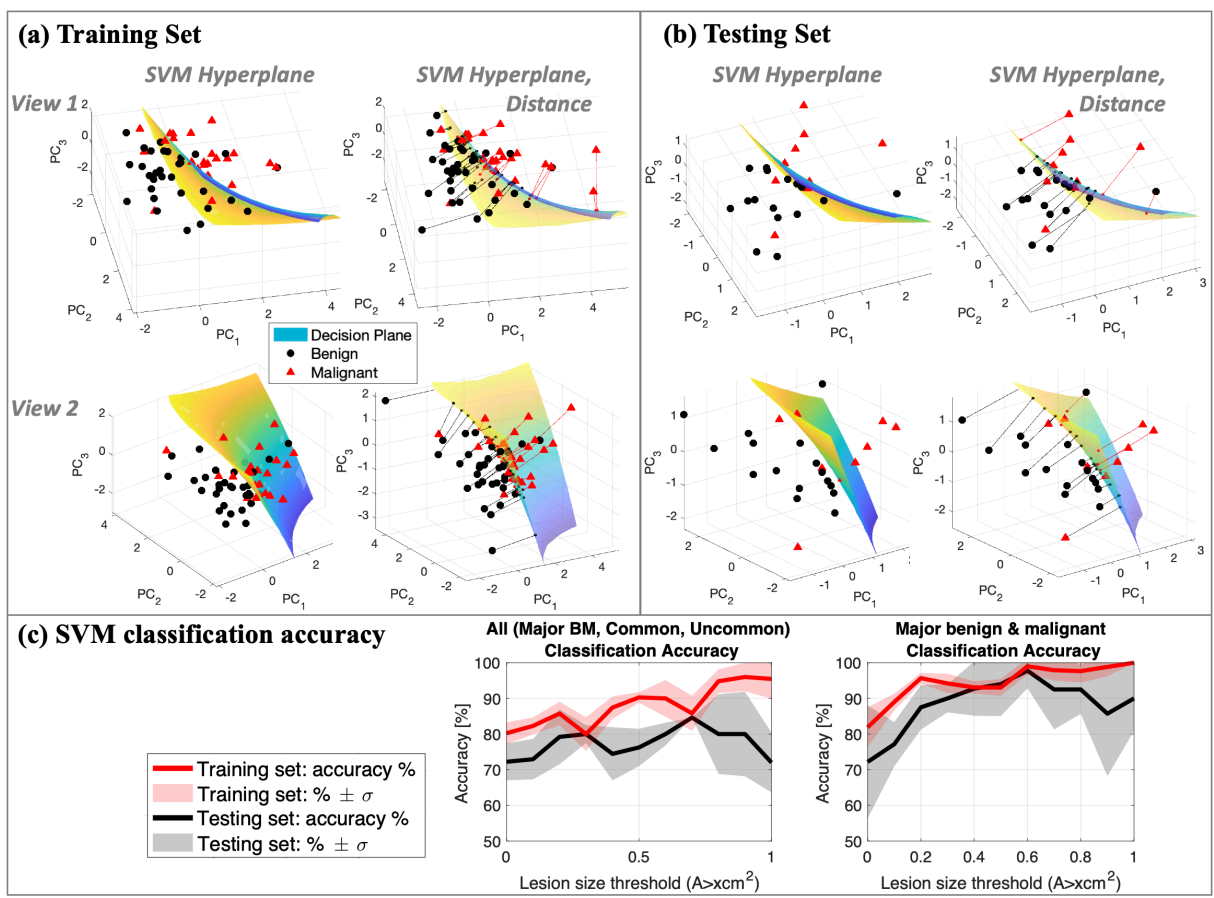

**Figure 9** SVM hyperplane and SVM distance lines for (a) a training set and (b) a testing set. Two different views are provided. (c) SVM classification accuracy differentiating benign and malignant cases. The left plot includes all categories of major, common, and uncommon cases. The right plot only includes major benign and malignant cases.

*C. DSI performance*

Fig. 10 shows the quantitative outputs of AUC, accuracy, sensitivity, and specificity. We compared results from the radiologists, S-detect, and our quantification using PC1, projection, and SVM distance. The results were plotted for training/testing sets and for major and all categories. Fig. 10 (a) and (b) include major categories of benign/malignant and all cases which includes major categories, common, and uncommon cases. For each plot, the performances were presented along with the lesion size thresholds ranging from 0 to 1 cm$^2$, showing lesion size dependency for breast tissue characterization. Each data point was obtained from the average of the 5 measurements since we generated 5 different sets of training and testing sets. Our quantification showed higher outputs (AUC, accuracy, sensitivity, and specificity) than the radiologists or S-detect. Further, among our 3 quantifications, the SVM distance parameter tends to show the best performance compared with PC1 and projection, but the results from PC1 and projection are still better than BI-RADS and S-detect. For smaller lesions, our results are slightly better or comparable to those of the radiologists and S-detect. However, in legion sizes greater than 0.5 cm$^2$, the difference between our results and S-detect is increasing because the deep learning performance tends to decrease with increasing lesion size. Since S-detect resized the input images to the same pixel number, the images of larger lesions have higher down-sampling rates, which might cause information loss from the sampling process and thus lower the performance.

When comparing the results of the training and testing sets, we can conclude that our DSI training is not overfitted because the testing results are comparable to the training results with only slight differences, and most of the outputs (AUC, accuracy, sensitivity, and specificity) from the testing set are over 0.8.

Fig. 10 (c) provides representative plots with ±σ error ranges using AUC scores. Although there are slight overlaps between the parameters, the error bounds are not completely overlapped, indicating meaningful differences between the different approaches. These plots can support that our DSI results are (1) better than or comparable to radiologists' BI-RADS scores and (2) better than the S-detect. Fig. 10 (c) only provides plots for AUC scores, but the other metrics of accuracy, sensitivity, and specificity showed comparable error ranges with the AUC results. To provide simple and easy to read plots, we have excluded the error shadings in Fig. 10 (a-b).

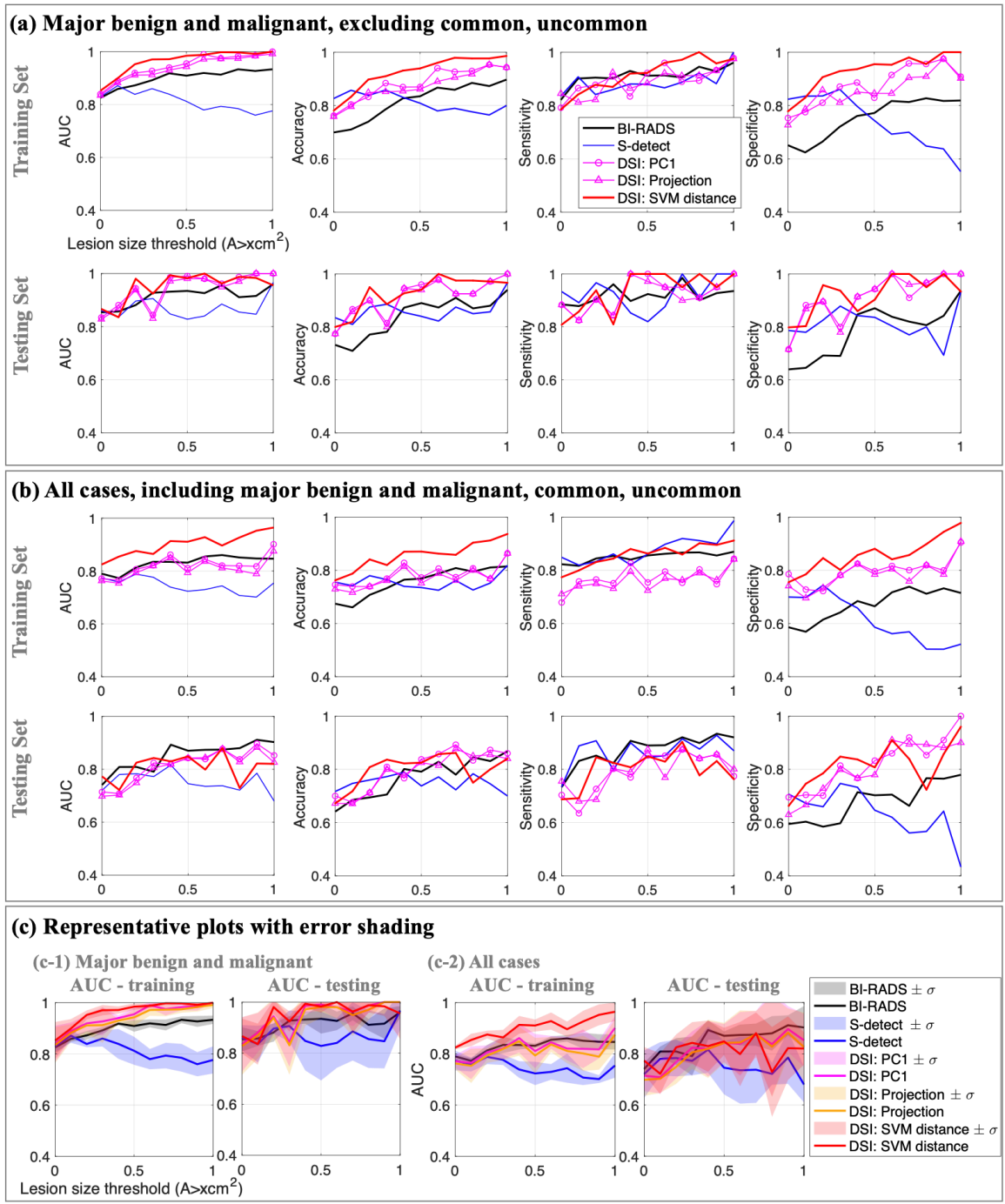

**Figure 10** Performance comparison: AUC, accuracy, sensitivity, and specificity from DSI (PC1, projection, and SVM distance), BI-RADS score from radiologists, and S-detect output utilizing a deep learning. The plots show quantification outputs of training and testing data sets. (a) includes only major categories of benign and malignant. (b) includes all categories of major, common, and not-common cases. (c) shows representative plots to show error bars using AUC scores. Measured average $\pm\sigma$ area was shaded for each output. We generated 5 independent sets of testing/training data, so each data point is average for the 5 measurements. The other results of accuracy, sensitivity, and specificity showed comparable error range to that of AUC.

Fig. 11 shows representative DSI images. The six patient cases in Fig. 11 have lower to higher BI-RADS scores in order: 3.20, 3.53, 3.90, 4.20, 4.47, and 4.93; the BI-RADS was scored by 10 radiologists and averaged. The localized probabilities of malignancy were estimated by PC1, projection, or SVM distance, which were encoded as colors between light blue to red, corresponding to benign to malignant, respectively, as provided by the color bar in Fig. 11. To obtain the images in Fig. 11, we applied trained DSI to major categories with lesion sizes greater than 0.4 cm2 and PC1 as color intensities. When calculating the combined parameters of PC1, projection, and SVM distance, we utilized the five features of H-scan color level, boundary shape parameter using convex hull, B-scan STD, B-scan boundary STD, and Burr b. H-scan color level is the only localized parameter, which may differentiate tissue types within a lesion. As post-processing for DSI display, 2D median and Gaussian filters were applied to the measured combined parameters (color intensities) within a lesion.

As shown in Fig. 11, the lower BI-RADS score cases have a higher proportion of green or yellow colors, but the higher BI-RADS score cases have a higher proportion of red color components, which demonstrates that DSI can differentiate BI-RADS score differences.

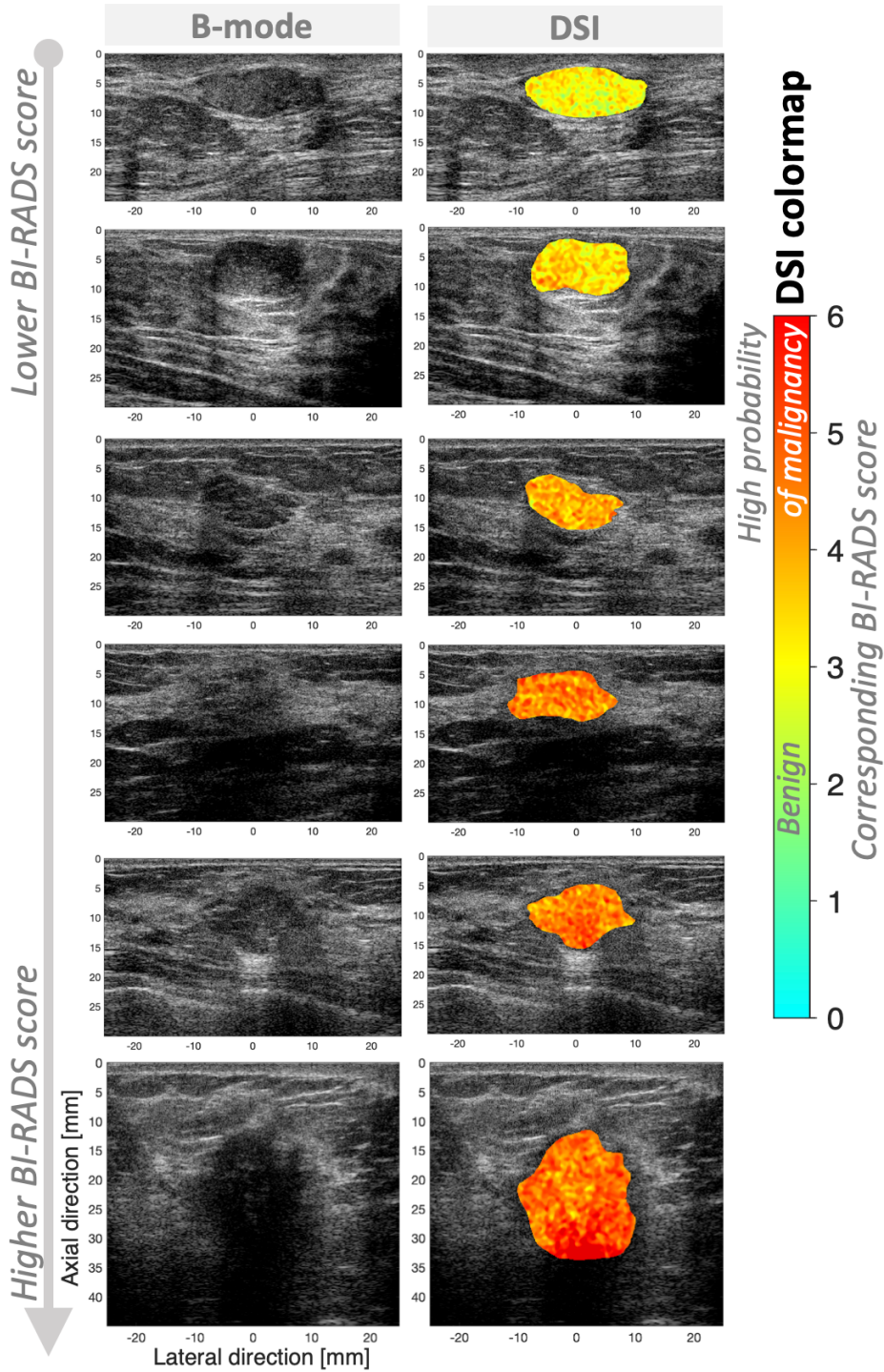

**Figure 11** DSI and its color bar. On B-mode images, the localized probability of malignancy was color-coded using the given color map from light blue to red, indicating higher probability of benign and malignant, respectively. From up to down in figure order, the lesions were scored by the 10 radiologists as BIRADS 3.20, 3.53, 3.90, 4.20, 4.47, and 4.93, respectively. For a quantitative guideline for DSI color levels and probability of malignancy, corresponding BI-RADS scores are provided. BI-RADS categories are defined by the American Cancer Society; 6: Biopsy-proven malignancy, 5: Highly suggestive of malignancy > 95%, 4: Suspicious for malignancy 2-94%, 3: Probably benign, malignancy<2%, 2: Benign, 1: Negative, 0: Incomplete.

The DSI colormap is provided in Fig. 11 with corresponding BI-RADS scores. The color levels are PC1, projection, or SVM distance, which visualize the probability of malignancy. Thus, to provide a quantitative guideline for breast cancer prediction, we suggest corresponding BI-RADS score for the DSI color levels based on correlation between BI-RADS and the color levels. An example mapping between BI-RADS and the color estimation is shown in Fig. 12, which was obtained using major categories with lesion sizes greater than 0.4 cm$^2$, consistent with the example images in Fig. 11. Our DSI training was performed using ground truth biopsy results, and BI-RADS scoring was also evaluated using the ground truth biopsy. As shown in Fig. 10, DSI outperformed BI-RADS, however their correlation can be used as the guideline with Spearman's correlation ($R_s$) coefficients of 0.81, 0.79, and 0.82 for PC1, projection, and SVM distance, respectively.

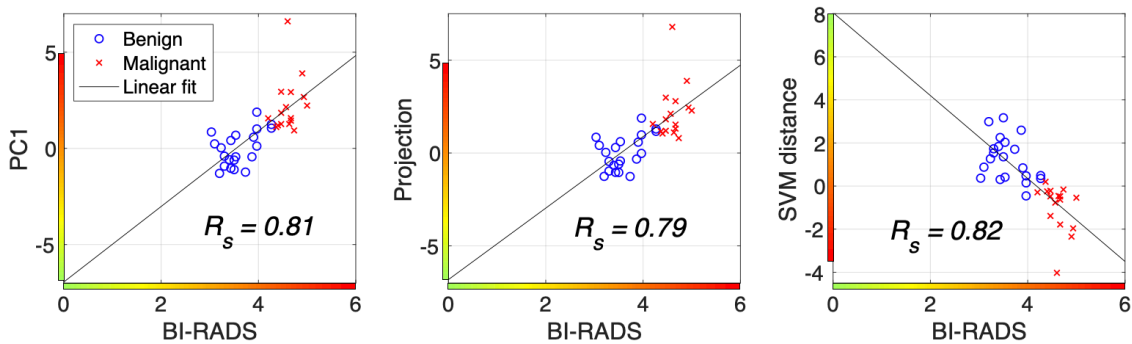

**Figure 12** An example correlation between BI-RADS score and probability of malignancy. Corresponding DSI colors are provided on x- and y-axis.

## V. DISCUSSION

Our DSI framework is capable of producing testing set classifications with AUC, accuracy, sensitivity, and specificity between 0.95 and 1.0 for lesion sizes greater than 0.6 cm$^2$ (0.9 to 1.0 for lesions greater than 0.4 cm$^2$) for the major categories of benign and malignant lesions. Results are less accurate for smaller and non-major lesion types, possibly due to the paucity of data in these cases. Furthermore, the DSI analysis leads to a visual representation with color overlay indicating the severity of the disease from detailed multiparametric analysis. Additionally, the DSI color levels can suggest a quantitative guideline for the probability of malignancy.

Our training only included 121 patients which may not be sufficient for comprehensive use, and thus we utilize the SVM classifier which advantageously requires less data compared with other deep learning approaches and simple classifiers. Moreover, we optimized the SVM parameters to avoid overfitting. Hence, we have constructed more robust planes, meaning the parameter settings were done to obtain smooth hyperplanes, although the planes still have higher-order curvature, as shown in Fig. 9 (a-b). However, as seen in Fig. 9 (c), when investigating all categories, the classification accuracy of the testing sets for lesion sizes greater than 0.8 $cm^2$ might be slightly lower than a result from successful training. This is because we included non-major categories and excluded smaller lesion sizes less than 0.8 $cm^2$, and thus the enrolled patients are less than 121 with non-major categories. Further studies which include sufficient patient numbers is required to construct more accurate hyperplanes that are more non-linear and sensitive to the training data set, and sufficiently accurate to quantify the probability of malignancy.

We demonstrated that our DSI framework yielded better performance than radiologists' BI-RADS scores and the established deep learning approach of S-detect (Fig. 10). First, for the comparison with the BI-RADS scores, radiologists utilized B-mode images, which may have limitations as the radiologists only rely on information from lesion boundary shapes and B-mode brightness. The post-processed B-mode images in ultrasound systems are optimized for the human eye, but in terms of signal processing, the images lose information while processing gray scale from channel RF data. Our DSI framework extracts features from summed RF data to utilize more information to characterize breast tissues, which cannot be seen by radiologists viewing only B-mode images. Secondly, for the comparison with S-detect, deep learning approaches generally utilize the post-processed B-mode images, reducing the computational complexity of utilizing raw data (such as RF) while reducing memory and processing time requirements. Ultrasound imaging applications in commercial ultrasound systems are sensitive to these issues. To address this within our DSI framework, we employed the SVM, which is less intensive in processing and therefore can utilize RF data parameters as its input. This approach allowed us to extract more information from ultrasound signals, which contributed to the better performance than S-detect.

This study demonstrated that our DSI approach has the potential to be used in breast cancer diagnosis. However, to be used as an application in clinical ultrasound systems, further studies to make it faster would be advantageous. This study has focused on obtaining better performance, but to consider tradeoffs between performance and time, the following need to be addressed. (1) In feature extraction, first for the H-scan parameter, we can use a smaller number of Gaussian-matched filters instead of the 256 filters used in this study. Reducing the filter number renders H-scan less precise but faster. Second, for boundary shape estimation, we performed estimation of a convex hull to compare a smooth shape and a real boundary shape. We also investigated ellipse fitting instead of estimation of the convex hull, but the fitting is more time-consuming and less accurate. However, it would be advantageous to find a more effective method to estimate a smooth and ideal shape (like ellipse or convex hull) for a lesion. (2) In the combined parameter for estimation of breast cancer progression, there are three parameters that have been considered as results of multiparametric analysis: PC1, projection, and SVM distance. According to our results, the SVM parameter performs the best, while PC1 and projection have comparable performance but slightly lower than the SVM parameter. However, comparing their processing times, PC1 is the fastest, comparable to projection, whereas SVM parameter estimation takes much more time than PC1 and projection. Thus, utilizing PC1 as the color intensity of DSI may result in slight performance loss, but could be much faster. (3) In the use of S-detect for lesion contour, S-detect is not working in real-time, and thus it outputs results onto a still frame. Our DSI utilizes the S-detect result of lesion boundary. Hence, the DSI would show results in a still frame, but efforts to make the overall boundary detection and DSI processing faster would be helpful for clinicians to see results in less time, including lesion identification and the color display.

## VI. CONCLUSION

We have developed a DSI framework for breast tissue characterization, including classification and prediction of the severity of breast cancer, utilizing multiparametric analysis. This study modified the previously proposed DSI classification of separate pathologies into a simpler classification of benign or malignant, and with a corresponding color overlay map to visualize the results of the analysis. This newer version of DSI enables the quantitative analysis of human breast lesions, including more complicated classes and types. The approach produces an AUC above 0.98 for the most common types of lesions and sizes.

Our approach to combine multiparametric analysis with a machine learning achieved better performance than both radiologists or the available deep learning system. Finally, we demonstrated that applying DSI not only allows for more accurate identification of breast lesions than deep learning alone, but also provides a visual display. Therefore, we anticipate that the DSI framework could be utilized to enable more accurate and rapid tissue characterization of human breast lesions in the clinical field.

*Acknowledgments*— This work was supported by National Institutes of Health grant R21EB025290. The authors are grateful to colleagues who conducted this breast study and provided RF data, including Zaegyoo Hah and Yeong-Kyeong Seong at Samsung Medison.